\documentclass[aps,a4paper,twocolumn,nofootinbib]{revtex4} 
\usepackage{mathptmx}
\usepackage{makeidx}
\usepackage{amsmath}
\usepackage{amsfonts}
\usepackage{amssymb}
\usepackage{mathrsfs}
\usepackage[mathscr]{euscript}
\usepackage[dvips]{graphicx}
\usepackage{fancyhdr}
\newcommand{\convol}{\star}
\newcommand{\grad}{\nabla}
\newcommand{\cross}{\times}

\renewcommand{\Vec}{\textbf}

\begin{document}

\title{A uni-directional optical pulse propagation equation for materials with both electric and magnetic responses}
\author{Paul Kinsler}
\email{Dr.Paul.Kinsler@physics.org}
\affiliation{
  Blackett Laboratory, Imperial College London,
  Prince Consort Road,
  London SW7 2AZ, 
  United Kingdom.
}

\begin{abstract}

I derive uni-directional wave equations for fields 
 propagating in materials with both electric and magnetic 
 dispersion and nonlinearity.
The derivation 
 imposes no conditions on the pulse profile 
 except that the material modulates the propagation 
 only slowly:
 i.e. that loss, 
 dispersion, and nonlinearity have only a small effect
 over the scale of a wavelength.
It also allows a direct term-to-term comparison
 of the exact bi-directional theory
 with its approximate uni-directional counterpart.

\end{abstract}

\lhead{\includegraphics[height=5mm,angle=0]{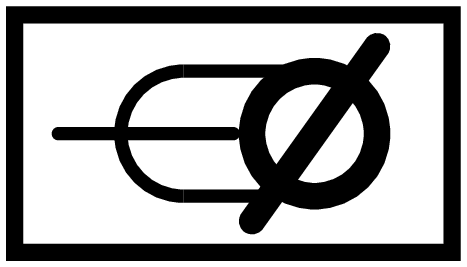}DBLNL-GPM}
\chead{UPE for [...] both  electric \& magnetic responses}
\rhead{
\href{mailto:Dr.Paul.Kinsler@physics.org}{Dr.Paul.Kinsler@physics.org}\\
\href{http://www.kinsler.org/physics/}{http://www.kinsler.org/physics/}
}

\date{\today}
\maketitle
\thispagestyle{fancy}

\section{Introduction}

In the past few years, 
 composite materials (``metamaterials'') 
 that demonstrate both an electric and magnetic response
 have been the subject of both experimental 
 and theoretical investigation.
Often the motivation for this research 
 is the potential for exotic applications:
 for example, superresolution \cite{Pendry-2000prl},
 or the possibility of ``trapped rainbow'' light storage
 \cite{Tsakmakidis-BH-2007n}.
Despite these interesting possibilities, 
 there is also the more basic need 
 for efficient methods for propagating
 optical pulses in such metamaterials,
 in particular in one-dimensional (1D) waveguide geometries.
Indeed, 
 methods for doing so have already led to interesting predictions
 (see, e.g., \cite{Scalora-SAPDMBZ-2005prl,Wen-XDTSF-2007pra,
                        DAguanno-MB-2008josab}).
However, 
 these methods, 
 and earlier ones, 
 (e.g. 
  \cite{Agrawal-NFO,Brabec-K-1997prl,Geissler-TSSKB-1999prl,Kinsler-N-2003pra})
 tend to rely on mechanisms 
 such as the introduction of a co-moving frame, 
 and assumptions that the pulse profile has negligible second-order
 temporal or spatial derivatives.
Assuming second-order derivatives are small may well be reasonable, 
 but it means that the pulse profile must remain well behaved.
This approximation therefore might well become poorly controlled
 \cite{Kinsler-N-2003pra,Kinsler-2002-fcpp}, 
 particularly for ultrashort or otherwise ultrawideband pulses, 
 or exotic or extreme material parameters.
Ideally we would prefer to make approximations based solely on the 
 material parameters of our device, 
 so as to avoid making assumptions about the state
 of an ever-changing propagating pulse.

Here I derive 1D wave equations for a waveguide with both 
 electric and magnetic dispersion, 
 and electric and magnetic nonlinearity.
I use the directional fields approach
 \cite{Kinsler-RN-2005pra,Kinsler-2007-envel}, 
 which allows us to directly write down a first-order wave equation 
 for pulse propagation without complicated derivation
 or approximation.
We simply look at the coupled forward and backward wave equations
 that are a direct re-expression
 of Maxwell's curl equations,
 and substitute in the appropriate dispersion and nonlinearity.
I also show separate examples for second- and third-order nonlinearities
 in both electric and magnetic responses, 
 although the effects can be combined if desired.
Note that these directional fields are applicable to more than just 
 pulse propagation, 
 as they have been used to simplify Poynting-vector-based
 approaches to electromagnetic continuity equations \cite{Kinsler-FM-2009ejp}.

The derivation makes only a single, 
 well-defined approximation to reduce the bi-directional
 forward-backward coupled model
 down to a single first-order wave equation --
 that of assuming small changes over the scale of a wavelength.
This approximation is remarkably robust for all physically realistic 
 parameter values -- 
 see \cite{Kinsler-2007josab} for an analysis
 focused on nonlinear effects; 
 more general considerations have been dealt with in terms of 
 factorized wave equations \cite{Kinsler-2008-fchhg}.
The resulting wave equation retains all the usual
 intuitive and analytical simplicity 
 of ordinary wave propagation equations, 
 unlike the computationally intensive approach of 
 a direct numerical solution of Maxwell's equations
 (see, e.g., \cite{Yee-1966tap,Flesch-PM-1996prl,
                 Gilles-MV-1999pre,Brabec-K-1997prl,
                 Tarasishin-MZ-2001oc,Tyrrell-KN-2005jmo}).

This paper is structured as follows:
 Directional fields and their 
 re-expression of the Maxwell curl equations
 is outlined in Section \ref{S-Gpm}, 
 followed by the reduction of the bi-directional wave equation 
 into a uni-directional form in Section \ref{S-waveeqns}.
Section \ref{S-chi3} shows
 wave equations for a doubly-nonlinear third-order nonlinearity material, 
 and Section \ref{S-chi2} does the same for a second-order case.
In Section \ref{S-discussion} propagation under the influence
 of typical metamaterial responses is discussed, 
 and I conclude in Section \ref{S-conclude}.

%
\section{Directional fields}\label{S-Gpm}

The directional fields approach \cite{Kinsler-RN-2005pra} 
 allows us to write down wave equations
 for hybrid electromagnetic fields $\Vec{G}^\pm$.
Note that here I define $\Vec{G}^\pm$ 
 with an alternate 
 (and more sensible) sign convention than previously.
Further, 
 I also allow for more general types of polarization and magnetization
 in such a way as to provide a \emph{simpler} presentation.
For propagation along the unit vector $\Vec{u}$, 
 the propagation (curl) equation for $\Vec{G}^\pm$ is written
 in the frequency domain
 as\footnote{See derivation in Appendix \ref{S-derive}.}
~
\begin{align}
  \grad \cross \Vec{G}^{\pm}
&=
 \pm
  \imath \omega
  \alpha_r \beta_r
  \Vec{u} \cross \Vec{G}^{\pm}
 ~
 \pm
  \imath
  \omega
  \beta_r
  \Vec{u} \cross \Vec{P}_c
 ~
 +
  \imath
  \omega
  \alpha_r
  \mu_0 \Vec{M}_c
\label{EQN-BASIC-DZGPM}
\end{align}
with
~
\begin{align}
  \Vec{G}^{\pm}(\omega)
&=
  \alpha_r(\omega)
  \Vec{E}(\omega)
 \pm
  \beta_r(\omega)
  \Vec{H}(\omega)
  \cross 
  \Vec{u}
\label{eqn-basic-Gpm}
.
\end{align}
Here the electric response of the material is encoded in two parts:
 a spatially invariant linear response component $\alpha_r$, 
 and the remaining contributions (of any type) in $\Vec{P}_c$.
Similarly, 
 the magnetic response is divided up in the same way 
 between $\beta_r$ and $\mu_0 \Vec{M}_c$.
Generally we will put the entire non-lossy linear response of the medium
 (i.e. the dispersion)
 into the reference parameters $\alpha_r$ and $\beta_r$, 
 although it may also be convenient to specify only that 
 the product $\alpha_r \beta_r$ is real
 (cf. \cite{Kinsler-2009pra}).
All the nonlinear responses and other complications (``corrections''), 
 such as spatial variations in the material parameters,
 remain in $\Vec{P}_c$ and $\Vec{M}_c$.
As an example, 
 in \cite{Kinsler-RN-2005pra} this approach was applied to 
 second-harmonic generation in a periodically poled dielectric crystal.
The time derivatives of these corrections $\Vec{P}_c$ and $\Vec{M}_c$
 correspond to bound 
 electric and magnetic currents respectively \cite{Kinsler-FM-2009ejp}.
These $\Vec{P}_c$ and $\Vec{M}_c$ are functions
 of both fields $\Vec{E}$ and $\Vec{H}$, 
 i.e. $\Vec{P}_c \equiv \Vec{P}_c(\Vec{E},\Vec{H})$
 and $\Vec{M}_c \equiv \Vec{M}_c(\Vec{E},\Vec{H})$.
If we choose instead to have frequency-independent $\alpha_r$ and $\beta_r$, 
 then the remaining linear response can simply be included
 in $\Vec{P}_c$ and $\Vec{M}_c$; 
 in this case the ``weak loss and nonlinearity'' condition
 I use later to decouple forward
 and backward fields would then need to be broadened to include
 weak dispersion as well
 (also see \cite{Kinsler-RN-2005pra} for more discussion).
However, 
 neither version imposes any requirements on the pulse profile.

It is useful to give a simple example of the directional fields
 to provide some insight into their nature.
In the pure transverse plane-polarized case,  
 with fields propagating along the $z$ direction, 
 and frequency-independent (material parameters) 
 permitivitty $\epsilon_r$
 and permeability $\mu_r$,
 we can write
~
\begin{align}
  G_x^\pm
&=
  \sqrt{\epsilon_r}
  E_x
 \pm
  \sqrt{\mu_r}
  H_y,
\\
  G_y^\pm
&=
  \sqrt{\epsilon_r}
  E_y
 \mp
  \sqrt{\mu_r}
  H_x
,
\label{eqn-eample-Gpm}
\end{align}
where this simple $G_x^\pm$ definition matches 
 the original proposal of Fleck \cite{Fleck-1970prb}.

It is worth considering how reflections arise
 in this picture based on spatially invariant reference parameters
 augmented by corrections terms.
Leaving aside for now the distinctions between spatially propagated fields
 and temporally propagated ones 
 (see the discussion in \cite{Kinsler-2008-fchhg}), 
 transition to a new media can be handled in two ways.
First, 
 we could map the existing fields $\Vec{G}^\pm$
 onto new ones $\Vec{G}_n^\pm$ based on new reference parameters
 $\alpha_{rn}$ and $\beta_{rn}$.
Here a pure $\Vec{G}^+$ field would separate into two pieces, 
 one a forward propagating $\Vec{G}_n^+$, 
 and the other a ``reflected'' backward propagating $\Vec{G}_n^-$.
Second, 
 we might retain the existing reference parameters, 
 and have modified corrections $\Vec{P}_c'$ and $\Vec{M}_c'$.
These altered correction terms then couple
 the forward and backward directed fields, 
 inducing the necessary reflection in $\Vec{G}^-$; 
 although as a side-effect of our now no longer optimal
 $\alpha_r$ and $\beta_r$, 
 the forward evolving field is made up of coupled $\Vec{G}^+$ and $\Vec{G}^-$
 components \cite{Kinsler-RN-2005pra}.

%
\subsection{Material response}

We define the electric and magnetic material response
 in the frequency domain, 
 as it greatly simplifies the description of the linear components.
Let us chose a \emph{reference} behaviour 
 given by $\epsilon_r(\omega), \mu_r(\omega)$,
 and use them to define reference parameters 
   $\alpha_r(\omega)= \sqrt{\epsilon_r(\omega)}$
 and $\beta_r(\omega) = \sqrt{\mu_r(\omega)}$.
Note that these are allowed to have 
 a frequency dependence \cite{Kinsler-RN-2005pra}, 
 and that $\alpha_r \beta_r$ is just the reciprocal
 of the (reference) speed of light in the medium
 (i.e. $n_r/c$).
We therefore have that the electric displacement and magnetic 
 fields are
~
\begin{align}
  \Vec{D}(\omega)
&=
  \epsilon_0 \Vec{E}(\omega) + \Vec{P}(\omega)
~
=&
  \epsilon_r(\omega) \Vec{E}(\omega) + \Vec{P}_c(\omega)
,
\\
  \Vec{B}(\omega)
&=
  \mu_0 \Vec{H}(\omega) + \Vec{M}(\omega)
=&
  \mu_r(\omega) \Vec{H}(\omega) + \mu_0 \Vec{M}_c(\omega)
.
\end{align}

To give a specific example, 
 we can define frequency-dependent 
 loss and dispersive corrections by $\kappa_\epsilon(\omega)$
 and $\kappa_\mu(\omega)$, 
 along with (e.g.) 
 independent third-order nonlinearities
 $\chi_\epsilon, \chi_\mu$ to both
 the material responses; 
 although any appropriate expression can be used --
 even magneto-electric or other types.
Thus we can write the frequency domain expressions
~
\begin{align}
  \Vec{P}_c (\omega)
&=
  \alpha_r^2
  \kappa_\epsilon
  \Vec{E}
 +
  \epsilon_0
  \mathscr{F}
  \left[
    \chi_\epsilon E^2(t)
  \right]
  \convol
  \Vec{E}
\label{eqn-components-epsilon-eg}
\\
  \mu_0
  \Vec{M}_c (\omega)
&=
  \beta_r^2 
  \kappa_\mu
  \Vec{H}
 +
  \mu_0
  \mathscr{F}
  \left[
    \chi_\mu H^2(t)
  \right]
  \convol
  \Vec{H}
,
\label{eqn-components-mu-eg}
\end{align}
 where $\mathscr{F}[...]$ takes the Fourier transform
 (which is necessary because nonlinear effects are defined
 in the time domain as powers of the field) 
 and $\convol$ denotes a convolution
 [i.e.,
 $a \convol b = \int a(\omega) b(\omega-\omega') d\omega'$].
If the nonlinearity is time dependent, 
 then the simple $\chi_\epsilon E^2$ type terms 
 can be replaced with the appropriate convolution.
In general, 
 it is best to pick $\alpha_r, \beta_r$ subject to the condition
 that the sizes of $\Vec{P}_c$ and $\Vec{M}_c$ 
 are minimised.

In a double-negative material 
 (with both $\epsilon, \mu < 0$)
 we would get imaginary $\alpha_r, \beta_r$, 
 changing the complex phase of $\Vec{G}^\pm(\omega)$
 away from that given by the original $\Vec{E}$ and $\Vec{H}$.
Since this is in the frequency domain, 
 it converts into a phase shift in the time domain, 
 so although imaginary-valued $\alpha_r, \beta_r$ might
 seem inconvenient, 
 it does not give unphysical results.

%
\section{Wave equations}\label{S-waveeqns}

Starting with the vectorial curl equation \eqref{EQN-BASIC-DZGPM}, 
 I first take the 1D,
 but bi-directional,
 limit, 
 and describe the approximation necessary 
 to produce a simpler uni-directional form.
After this, 
 I discuss how the common transformations
 used in optical wave equations can be applied in this context.
All equations and field quantities are in the frequency ($\omega$) domain, 
 unless explicitly noted otherwise.

%
\subsection{Bi-directional wave equations}

Here we set $\Vec{u}$ along the $z$ axis without loss of generality, 
 and consider just an $x$ polarized wave 
 (i.e., consisting of $E_x, H_y$).
This means we use the $y$ component of eqn. \eqref{EQN-BASIC-DZGPM}
 with $\partial_z = d/dz$, 
 so that the wave equations for the 
 \emph{full spectrum} fields ${G}_x^\pm(\omega)$,
 coupled by corrections $P_x \equiv {P}_{c,x}(\omega)$ 
 and $M_y \equiv {M}_{c,y}(\omega)$
 are
~
\begin{align}
  \partial_z {G}_x^{\pm}
&=
 \pm
  \imath \omega
  \alpha_r \beta_r
  {G}_x^{\pm}
 \quad
 \pm
  \imath
  \omega
  \beta_r
  {P}_x
 \quad
 +
  \imath
  \omega
  \alpha_r
  {M}_y
.
\label{eqn-1d-dzGxpm-raw}
\end{align}
Following the detailed discussion
 in \cite{Kinsler-2008-fchhg}\footnote{Section III}, 
 we say that this wave equation \emph{propagates} (``steps'') the fields
 forward along the $z$ direction using 
 oppositely \emph{directed} fields ${G}_x^+$ and ${G}_x^-$.
These fields can be written as functions of either time or frequency, 
 and pulses they describe therefore \emph{evolve} (travel) 
 forward or backward in time.

Consider the example case with parameters 
 $\kappa_\epsilon(\omega)$ and $\kappa_\mu(\omega)$, 
 and $\chi_\epsilon, \chi_\mu$
 in eqns. \eqref{eqn-components-epsilon-eg} and 
 \eqref{eqn-components-mu-eg}.
Defining $k_r(\omega) = \omega \alpha_r(\omega) \beta_r(\omega)$, 
 we get
~
\begin{align}
  \partial_z {G}_x^{\pm}
&=
 \pm
  \imath k_r
  {G}_x^{\pm}
 ~~
 \pm
  \frac{\imath k_r \kappa_\epsilon }{2}
  \left(
    {G}_x^{+}
   +
    {G}_x^{-}
  \right)
 +
  \frac{\imath k_r \kappa_\mu }{2}
  \left(
    {G}_x^{+}
   -
    {G}_x^{-}
  \right)
\nonumber
\\
& \qquad 
 \pm
  \frac{\imath k_r}{2}
    \frac{\epsilon_0}{\epsilon_r}
    \mathscr{F}\left[ \chi_\epsilon E^2 \right]
  \convol
  \left(
    {G}_x^{+}
   +
    {G}_x^{-}
  \right)
\nonumber
\\
& \qquad \quad
 +
  \frac{\imath k_r}{2}
    \frac{\mu_0}{\mu_r}
    \mathscr{F}\left[ \chi_\mu H^2 \right]
  \convol
  \left(
    {G}_x^{+}
   -
    {G}_x^{-}
  \right)
.
\label{eqn-1d-dzGxpm}
\end{align}
Note that even for a frequency-independent choice
 of the reference parameters $\alpha_r$ and $\beta_r$, 
 the reference wave vector $k_r$ retains
 a (linear) frequency dependence.
Also, 
 the dispersion and/or loss parameters $\kappa_\epsilon, \kappa_\mu$
 are directly related to $\epsilon$ and $\mu$ respectively, 
 and not to a refractive index $n$ or wavevector $k$.
This is why there is a factor of $1/2$ associated with their 
 appearance in eqn. \eqref{eqn-1d-dzGxpm} and subsequently.

If written in the time domain, 
 these wave equations are seen to propagate the full temporal history
 of a field forward in space.
There, 
 the reference propagation given by $\pm \imath k_r G_x^\pm$
 becomes a convolution if $k_r$ retains a nontrivial frequency dependence.
However, 
 if we expand $k_r(\omega)$ around a central frequency $\omega_1$ 
 in powers of $\omega-\omega_1$, 
 we can instead convert it (in the time domain) 
 into a Taylor series in time derivatives, 
 which is a popular alternative to the frequency domain form used here.
However, 
 if implementing a split-step Fourier method of solving these wave equations,
 dispersion is applied in the frequency domain, 
 so that in general such an expansion is an unnecessary complication.

%
\subsection{Uni-directional approximation}

Now we apply the approximation:
 that the effect of any correction terms is small 
 over propagation distances of one wavelength --
 or, 
 if you prefer, 
 over time intervals of one optical period.
This translates into a weak loss and nonlinearity assumption; 
 and if the correction terms $\Vec{P}_c$ and $\Vec{M}_c$ include dispersion, 
 a weak dispersion assumption is also made.
These are rarely very stringent approximations.
If $|\Vec{P}_c| << |\Vec{D}|$
 and $|\mu_0\Vec{M}_c| << |\Vec{B}|$,
 then a forward $\Vec{G}^{+}$ 
 has minimal co-propagating $\Vec{G}^{-}$ \cite{Kinsler-RN-2005pra}.
Further, 
 the forward field has a wave vector $k_r$ 
 evolving as $\exp(+\imath k_r z)$, 
 but any generated backward component will evolve as $\exp(-\imath k_r z)$.
This gives a very rapid relative oscillation $\exp(-2\imath k_r z)$, 
 which will quickly average to zero.
Nevertheless, 
 although achievable optical nonlinearity coefficients 
 fall well within this approximation, 
 care may need to be taken with the dispersion, 
 particularly if near a band edge or in the vicinity
 of a narrow resonance.

A directly comparable approximation is treated exhaustively
 in \cite{Kinsler-2008-fchhg},
 where although applied to bi-directional factorizations 
 of the second-order wave equation, 
 the physical considerations are exactly the same:
 Deviations from the reference behaviour 
 over a propagation distance of one wavelength should be small.
Note that the slow evolution approximation applied here
 is not the same as other ``slowly varying'' types of approximation 
 [e.g., the slowly varying envelope approximation (SVEA)] --
 although the physical motivation is similar, 
 the approach used here is far less restrictive.

After we apply this weak correction or ``slow evolution'' approximation, 
 we set the initial value of ${G}_x^{-} \equiv 0$, 
 and can be sure that it will stay negligible.
Thus eqn. \eqref{eqn-1d-dzGxpm-raw} 
 for the full spectrum, forward directed field ${G}_x^{+}(\omega)$
 can be written as
~
\begin{align}
  \partial_z {G}_x^{+}
&\simeq
 +
  \imath \omega
  \alpha_r \beta_r
  {G}_x^{+}
 ~~
 +
  \imath
  \omega
  \beta_r
  {P}_x
 ~~
 +
  \imath
  \omega
  \alpha_r
  \mu_0 {M}_y
.
\label{eqn-unidir-Gp}
\end{align}
Alternatively, 
 we can scale the ${G}_x^{+}$ field
 so that it has the same units and scaling as the electric field, 
 using ${F}_x^{+}(\omega) = {G}_x^{+}(\omega) / 2 \alpha_r(\omega)$.
This gives
~
\begin{align}
  \partial_z {F}_x^{+}
&\simeq
 +
  \imath \omega
  \alpha_r \beta_r
  {F}_x^{+}
 ~~
 +
  \imath
  \frac{\omega\beta_r}{2\alpha_r}
  {P}_x^+
 ~~
 +
  \imath
  \frac{\omega}{2}
  \mu_0
  {M}_y^+
\label{eqn-unidir-Fp}
.
\end{align}
Note that
 $E_x^+={G}_x^{+} / 2 \alpha_r \equiv {F}_x^{+}$
 and $H_y^+={G}_x^{+} / 2 \beta_r \equiv {F}_x^{+} \alpha_r/\beta_r$, 
 since ${G}_x^{-}=0$.
In either version of these uni-directional equations, 
 $P_x^+ \equiv P(E_x^+, H_y^+)$
 and 
 $M_y^+ \equiv M(H_y^+, E_x^+)$ --
 the uni-directional (residual) polarization ${P}_x^+$
 and (residual) magnetization ${M}_y^+$
 should not be written as functions of the total fields ${E}_x$ and ${H}_y$.

%
\subsection{Modifications}
\label{S-modifications}

Either of eqns. \eqref{eqn-unidir-Gp} or \eqref{eqn-unidir-Fp}
 by themselves are sufficient to model the propagation of 
 the electric and magnetic fields.
However, 
 there are many traditional simplifications which can be applied, 
 and which in other treatments
 are even sometimes \emph{required}
 in order obtain a simple evolution equation.
In particular, 
 the various envelope equations 
 \cite{Brabec-K-1997prl,Kinsler-N-2003pra,Geissler-TSSKB-1999prl,
         Scalora-C-1994oc}
 all use co-moving and/or envelopes
 as a preparation for discarding inconvenient derivatives:
 Here such steps are optional extras.

These are all considered in more detail
 for a factorised wave equation approach
 in \cite{Kinsler-2008-fchhg}, 
 but here I have adapted them for this context.

\begin{enumerate}

\item
A \emph{co-moving frame} can now be added, 
 using $t'=t-z/v_f$.
This is a simple linear process that causes no extra complications;
 the leading right-hand side (RHS)
 $\imath \alpha_r \beta_r \omega = \imath k_r$ term
 is replaced
 by $\imath (\alpha_r \beta_r \omega \mp k_f)$,
 for frame speed $v_f = \omega_0/k_f$.
Setting $k_f=k_r(\omega_0)$ will cancel the phase velocity $v_p$
 of the pulse at $\omega_0$, 
 not the group velocity.

\item
The \emph{field can be split up} into pieces localized
 at certain frequencies, 
 as done in descriptions of optical parametric amplifiers or Raman combs
 (as in, e.g., \cite{Kinsler-N-2003pra,Kinsler-N-2004pra,Kinsler-N-2005pra}).
 The wave equation can then be separated into one equation for 
 each piece, 
 coupled by the appropriate frequency-matched polarization terms
 (see, e.g., \cite{Kinsler-NT-2006-phnlo}).

\item
A \emph{carrier-envelope} description of the field 
 can easily be implemented
 with the usual prescription of \cite{Gabor-1946jiee,Boyd-NLO}
 ${F_x^+}(t) = A(t) \exp [\imath (\omega_1 t - k_1 z)] 
        + A^*(t) \exp [-\imath (\omega_1 t - k_1 z)] $
 defining an envelope $A(t)$ 
 with respect to carrier frequency $\omega_1$ and wave vector $k_1$;
 this also provides a built-in a co-moving frame
 $v_f = \omega_1/k_1$.
Multiple envelopes centred at different carrier frequencies
 and wave vectors ($\omega_i$, $k_i$) can also be used
 \cite{Kinsler-NT-2006-phnlo,Boyd-NLO}.

\item
\emph{Bandwidth restrictions} might be added (see below), 
 either to ensure
 a smooth envelope or to simplify the wave equations; 
 in addition they might be used to separate out or neglect 
 frequency mixing terms or harmonic generation.
As it stands, 
 no bandwidth restrictions were applied when deriving 
 eqns. \eqref{eqn-unidir-Gp} or \eqref{eqn-unidir-Fp} --
 there are only the limitations introduced 
 by the dispersion and/or polarization models to consider.
Typically we would expand the model parameters to the first few orders
 about some convenient reference frequency $\omega_0$.

\item
\emph{Mode averaging} is where the transverse extent
 of a propagating beam is not explicitly modeled, 
 but is subsumed into a description of a transverse mode profile; 
 as such it is typically applied to situations involving 
 optical fibres or other waveguides.
Thus we could use mode averaging
 when calculating the effective dispersion or nonlinear parameters.
See,
 for example,
 \cite{Laegsgaard-2007oe} for a recent approach, 
 which goes beyond a simple addition of a frequency dependence
 to the ``effective area'' of the mode, 
 and generalizes the effective area concept itself.

\end{enumerate}

%
\subsection{Diffraction}\label{S-diffraction}

One important feature lacking in this approach is the handling
 of transverse effects such as diffraction, 
 although they can be inserted by hand
 (at least in the paraxial limit)
 by adding the term $\imath (\partial_x^2 + \partial_y^2)F_x^+/2k_r$
 to the RHS \cite{Kinsler-2007-envel}.
However, 
 no treatment of transverse effects 
 has been achieved in a native directional fields description
 on the basis of the first-order equations -- 
 although transverse terms arise naturally in the second-order equation 
 resulting from taking the curl of eqn. \eqref{EQN-BASIC-DZGPM}.
Treating nonlinear diffraction \cite{Boardman-MPS-2000oqe}
 suffers the same difficulties, 
 although presumably it might be incorporated in an analogous
 way as to ordinary diffraction.

%
\section{Third-order nonlinearity}\label{S-chi3}

Third-order nonlinearities are common in many materials, 
 for example in the silica used to make optical fibres
 (see, e.g., \cite{Agrawal-NFO}).
There are many applications of significant scientific interest, 
 for example,
 white light supercontinnua 
 \cite{Alfano-S-1970prl,Alfano-S-1970prl-2,Dudley-GC-2006rmp},
 optical rogue waves \cite{Solli-RKJ-2007n}; 
 or filamentation \cite{Braun-KLDSM-1995ol,Berge-S-2009dcdsa}.

Here we study propagation through such a material, 
 with non-reference linear responses 
 $\kappa_\epsilon, \kappa_\mu$ (describing e.g. loss and dispersion), 
 and instantaneous magnetic third-order nonlinearity $\chi_\mu$
 \cite{Klein-WFL-2007oe} 
 along with the more common electric type 
 $\chi_\epsilon$.
For such a system, 
 and for plane-polarized fields, 
 the propagation equation is
~
\begin{align}
  \partial_z {F}_x^{+}
&=
 +
  \imath
  k_r 
  \left[
    1
   +
    \frac{\kappa_\epsilon}{2}
   +
    \frac{\kappa_\mu}{2}
  \right]
  {F}_x^{+}
\nonumber\\
& \qquad
 ~
 +
  \frac{\imath k_r}{2}
  \frac{\epsilon_0}{\epsilon_r}
  \left[
    \chi_\epsilon
   +
    \frac{\mu_0}{\epsilon_0}
    \frac{\epsilon_r^2}{\mu_r^2}
    \chi_\mu
  \right]
   \mathscr{F}\left[{F}_x^{+2}(t)\right]
   \convol
    {F}_x^{+}
.
\label{eqn-1d-dzFxpm}
\end{align}
This is a generalized nonlinear Schr\"odinger (NLS) equation, 
 and it retains both the full field 
 (i.e. uses no envelope description)
 and the full nonlinearity 
 (i.e. includes third-harmonic generation).
The only assumptions made are that of
 transverse fields and weak dispersive and nonlinear responses; 
 these latter assumptions allow us to decouple
 the forward and backward wave equations.
This decoupling  allows us, 
 without any extra approximation,
 to reduce our description to one of forward-only pulse propagation.
The specific example chosen here is for a cubic nonlinearity, 
 but it is easily generalized to the noninstantaneous case
 or even other scalar nonlinearities.

We can transform eqn. \eqref{eqn-1d-dzFxpm}
 into one closer to the ordinary NLS equation by 
 representing the field in terms of an envelope and carrier
~
\begin{align}
  {F}_x^+(t) 
&= A(t) \exp \left[\imath \left(\omega_0 t - k_0 z\right)\right]
        + A^*(t) \exp \left[-\imath \left(\omega_0 t - k_0 z\right)\right] 
,
\label{eqn-eg-envcar}
\end{align}
 where we choose the carrier wave vector
 to be $k_0=k_r(\omega_0)=\omega_0 \alpha_r(\omega_0) \beta_r(\omega_0)$. 
After separating into a pair of complex-conjugate equations
 (one for $A$ and one for $A^*$),
 and ignoring the off-resonant third-harmonic generation term,
 this gives us the expected NLS equation
 without diffraction. 
The chosen carrier effectively moves us into a frame
 that freezes the carrier oscillations, 
 but this phase velocity ($v_p=\omega/k$) frame 
 differs from one that is co-moving with the pulse envelope
 (i.e., one moving at the group velocity $v_g=\partial\omega/\partial k$).
After we transform into a frame co-moving with the group velocity
 at $\omega_0$, 
 where $\Delta_g=\omega_0 [v_g^{-1}(\omega_0)-v_p^{-1}(\omega_0)]$,
 the wave equation for ${A}(\omega)$ is 
~
\begin{align}
  \partial_z {A}
&=
 +
  \imath 
  K(\omega)
  {A}
 ~~
 +
  \frac{\imath k_r}{2}
  \frac{\epsilon_0}{\epsilon_r}
    \mathscr{F}
    \left[
      2
      \chi
      {\left| A(t) \right|^2}
      {A}(t)
    \right]
,
\label{eqn-eg-nls}
\end{align}
 where $K(\omega) = k_r [\kappa_\epsilon(\omega) + \kappa_\mu(\omega)]/2
                     + \Delta_g$; 
 and $\chi = \chi_\epsilon - (\mu_0 \epsilon_r^2/\epsilon_0 \mu_r^2) \chi_\mu$.
All that has been assumed to derive this standard envelope NLS equation
 is uni-directional propagation 
 and negligible third-harmonic generation.
The self-steepening term, 
 often seen in (or added to) NLS equations
 arises from the frequency dependence of $k_r$.
This self-steepening has both electric and magnetic contributions,
 which can be adjusted independently, 
 as has been pointed out by Wen \emph{et al.} \cite{Wen-XDTSF-2007pra}
 for the case of the SVEA limit.
In Section \ref{S-discussion},
 I discuss how the importance of each contribution
 varies with frequency for both a double-plasmon model 
 (as in \cite{Wen-XDTSF-2007pra}),
 and a wire-array and split-ring model more typical of practical metamaterials.

It is worth comparing this eqn. \eqref{eqn-eg-nls}
 to D'Aguanno \emph{et al.}'s \cite{DAguanno-MB-2008josab}
 eqn. (5) [hereafter eqn. (DMB5)].
Although in many respects they appear to be the same, 
 mine is far more general and can be applied (at least in principle)
 to an arbitrarily wide pulse bandwidth, 
 whereas theirs is subject to the rather restrictive SVEA.
For example, 
 my eqn. \eqref{eqn-eg-nls} results from 
 only one ``slow evolution'' approximation, 
 as opposed to the numerous steps, 
 substitutions, 
 and approximations in Section 2 of \cite{DAguanno-MB-2008josab}.
I also retain the possibility of arbitrary dispersion $K(\omega)$, 
 whereas theirs retains only the second-order part
 (i.e. as $\propto \partial_t^2$, 
 which in the frequency domain would be $\propto \omega^2$).
Indeed, 
 with the dispersion and nonlinear factors in my eqn. \eqref{eqn-1d-dzFxpm}
 combined, 
 that full-spectrum wave propagation equation is scarecly
 more complicated than eqn. (DMB5).
Similar remarks also hold when comparing eqn. \eqref{eqn-eg-nls} 
 to Wen \emph{et al.}'s \cite{Wen-XDTSF-2007pra}:
 but although Wen \emph{et al.}'s result is also restricted by the SVEA, 
 it does at least allow for diffraction.
Both, 
 however, 
 along with Scalora \emph{et al.}'s form \cite{Scalora-SAPDMBZ-2005prl}, 
 cannot model the full non-envelope field, 
 nor revert to an exact and explicitly bi-directional form, 
 as in my eqn. \eqref{eqn-1d-dzGxpm-raw} or \eqref{eqn-1d-dzGxpm}.

%
\section{Second-order nonlinearity}\label{S-chi2}

Treating a second-order nonlinearity 
 is more complicated than the third-order case, 
 since it typically couples the two possible polarization states
 of the field together.
Such interactions occur in materials 
 used for optical parametric amplification,
 and have long been used for a wide variety of applications
 (see, e.g., \cite{Boyd-NLO,OPOA-1993josab,Danielius-PSBDR-1992josab}).
To model the cross-coupling between the orthogonally polarised fields,
 it is necessary to solve for both field polarizations; 
 and to allow for the birefringence we need two pairs of 
 (non-reference)
 linear responses, 
 i.e. $\kappa_{\epsilon x}, \kappa_{\epsilon y}$
 and $\kappa_{\mu x}, \kappa_{\mu y}$.

As an example, 
 I choose a magnetic nonlinearity that couples $H_y$ and $H_x$
 in the same way as the electric nonlinearity couples $E_x$ and $E_y$, 
 although other configurations are possible.
This means that the $\beta_r\Vec{u}\cross\Vec{P}$ term
 in eqn. \eqref{EQN-BASIC-DZGPM},
 which represents the non-reference part of the electric response,
 needs to include those for the standard 
 second-order nonlinear terms
 (here $P_x \propto E_x E_y$ and $P_y \propto E_x^2$).
Similarly,
 the $\alpha_r \mu_0 \Vec{M}$ term has ones for the 
 complementary second-order nonlinear magnetic response.
Note that second-order nonlinear magnetic effects have been measured
 in split ring resonators
 by Klein \emph{et al.} \cite{Klein-EWL-2006s,Klein-WFL-2007oe}.

Since it is convenient, 
 I split the vector form of the $\Vec{G}^\pm$ wave equation 
 up into its transverse $x$ and $y$ components.
By noting that the definition of $G^\pm_y$
 means that $H_x^+ = -F_y^+ \alpha_r/\beta_r$, 
 the 1D wave equations can be written as
~
\begin{align}
  \partial_z {F}_x^{+}
&= 
 +
  \imath k_r
  \left[
    1
   +
    \frac{\kappa_{\epsilon x}}{2}
   +
    \frac{\kappa_{\mu y}}{2}
  \right]
  {F}_x^{+}
\nonumber\\
& \qquad +
  2
  \frac{\imath k_r}{2}
  \frac{\epsilon_0}{\epsilon_r}
  \left[
    \chi_\epsilon
   -
    \frac{\mu_0}{\epsilon_0}
    \left(\frac{\epsilon_r}{\mu_r}\right)^{\frac{3}{2}}
    \chi_\mu
  \right]
  \mathscr{F}
    \left[{F}_y^{+}(t) {F}_x^{+}(t)\right]
\label{eqn-chi2-x}
\end{align}\begin{align}
  \partial_z {F}_y^{+}
&= 
 +
  \imath k_r
  \left[
    1
   +
    \frac{\kappa_{\epsilon y}}{2}
   +
    \frac{\kappa_{\mu x}}{2}
  \right]
  {F}_y^{+}
\nonumber\\
& \qquad 
 ~+
  \frac{\imath k_r}{2}
  \frac{\epsilon_0}{\epsilon_r}
  \left[
    \chi_\epsilon
   +
    \frac{\mu_0}{\epsilon_0}
    \left(\frac{\epsilon_r}{\mu_r}\right)^{\frac{3}{2}}
    \chi_\mu
  \right]
  \mathscr{F}
    \left[{F}_x^{+}(t)^2\right]
\label{eqn-chi2-y}
.
\end{align}
These wave equations for the field
 are strikingly similar to the usual SVEA
 equations used to propagate narrowband pulses; 
 the main differences are the addition of terms
 for magnetic dispersion ($\kappa_{\mu x}, \kappa_{\mu y}$)
 and nonlinearity ($\chi_{\mu}$), 
 and the lack of a co-moving frame.

We can transform eqns. \eqref{eqn-chi2-x} and \eqref{eqn-chi2-y}
 into a form close to the usual equations for a parametric amplifier
 by representing the $x$ and $y$ polarized fields 
 in terms of three envelope and carrier pairs:
~
\begin{align}
  {F}_x^+(t) 
&= A_1(t) \exp \left[\imath \left(\omega_1 t - k_1 z\right)\right]
        + A_1^*(t) \exp \left[-\imath \left(\omega_1 t - k_1 z\right)\right]
\label{eqn-eg-envcar12}
\nonumber
\\
& \quad 
 +
  A_2(t) \exp \left[\imath \left(\omega_2 t - k_2 z\right)\right]
        + A_2^*(t) \exp \left[-\imath \left(\omega_2 t - k_2 z\right)\right]
\\
  {F}^+_y(t) 
&= A_3(t) \exp \left[\imath \left(\omega_3 t - k_3 z\right)\right]
        + A_3^*(t) \exp \left[-\imath \left(\omega_3 t - k_3 z\right)\right] 
,
\label{eqn-eg-envcar3}
\end{align}
 where $\omega_3 = \omega_1 + \omega_2$.
After separating into pairs of complex-conjugate equations
 (one each for all $A_i$ and $A_i^*$), 
 and ignoring the off-resonant polarization terms,
 we transform into a frame co-moving with the group velocity,
 although here we select the group velocity of a preferred frequency component,
 with $\Delta_{g}=\omega (v_{g}^{-1}-v_p^{-1})$.
The wave equations for the ${A}_i(\omega)$ are then
~
\begin{align}
  \partial_z {A}_1
&= 
 +
  \imath 
    K_1(\omega)
  {A}_1
 ~~
 +
  \frac{\imath k_0^2}
       {2 k_1}
    \chi^{-}
  \mathscr{F}
  \left[
    2 {A}_3(t) {A}_2^*(t)
  \right]
    e^{-\imath \Delta k z}
\label{eqn-chi2-A1}
\\
  \partial_z {A}_2
&= 
 +
  \imath 
    K_2(\omega)
  {A}_2
 ~~
 +
  \frac{\imath k_0^2}
       {2 k_2}
    \chi^{-}
  \mathscr{F}
  \left[
    2 {A}_3(t) {A}_1^*(t)
  \right]
    e^{-\imath \Delta k z}
\label{eqn-chi2-A2}
\\
  \partial_z {A}_3
&= 
 +
  \imath 
    K_3(\omega)
  {A}_3
 ~~
 +
  \frac{\imath k_0^2}
       {2 k_3}
     \chi^{+}
  \mathscr{F}
  \left[
    {A}_1(t) {A}_2(t)
  \right]
    e^{+\imath \Delta k z}
.
\label{eqn-chi2-A3}
\end{align}
Here 
 $K_{1,2}(\omega) = k_{1,2} [\kappa_{\epsilon x}(\omega) 
    + \kappa_{\mu y}(\omega)]/2 
    + \Delta_g$
 and $K_{3}(\omega) = k_{3} [\kappa_{\epsilon y}(\omega) 
    + \kappa_{\mu x}(\omega)]/2 
    + \Delta_g$;
 we choose $k_r$ for each equation differently
 (i.e., with $k_r \in \{k_1, k_2, k_3\}$); 
 also the phase mismatch term is $\Delta k = k_3 - k_2 - k_1$.
The combined nonlinear coefficient 
 is $\chi^\pm = \chi_\epsilon 
                \pm (\mu_0/\epsilon_0) (\epsilon_r/\mu_r)^{3/2} \chi_\mu$.

%
\section{Discussion}\label{S-discussion}

Examining the respective roles
 of the reference permittivity $\epsilon_r$ and 
 permeability $\mu_r$  in eqns. \eqref{eqn-1d-dzFxpm}
 and \eqref{eqn-chi2-x}, \eqref{eqn-chi2-y}, 
 we see that as far as dispersion and other linear effects
 are concerned, 
 the two components simply add.
In contrast, 
 their effect on nonlinear terms is more dramatic:
 with the ratio $Y^2=\epsilon_r/\mu_r$ 
 scaling the nonlinear corrections to the magnetization
 into the electric field units of $\Vec{F}^\pm$.
This is because $Y^2$ determines how much
 of a given directional field $\Vec{F}^\pm$
 is electric field and how much magnetic field; 
 large values of $Y^2$ correspond to cases where the magnetic field
 is most prominent.
Indeed, 
 $Y$ is just the reciprocal
 of the electromagnetic impedance of our chosen reference medium, 
 and only if $Y$ is real-valued do propagating fields exist, 
 since otherwise the fields become evanescent.

\begin{figure}
 \includegraphics[width=0.70\columnwidth,angle=0]{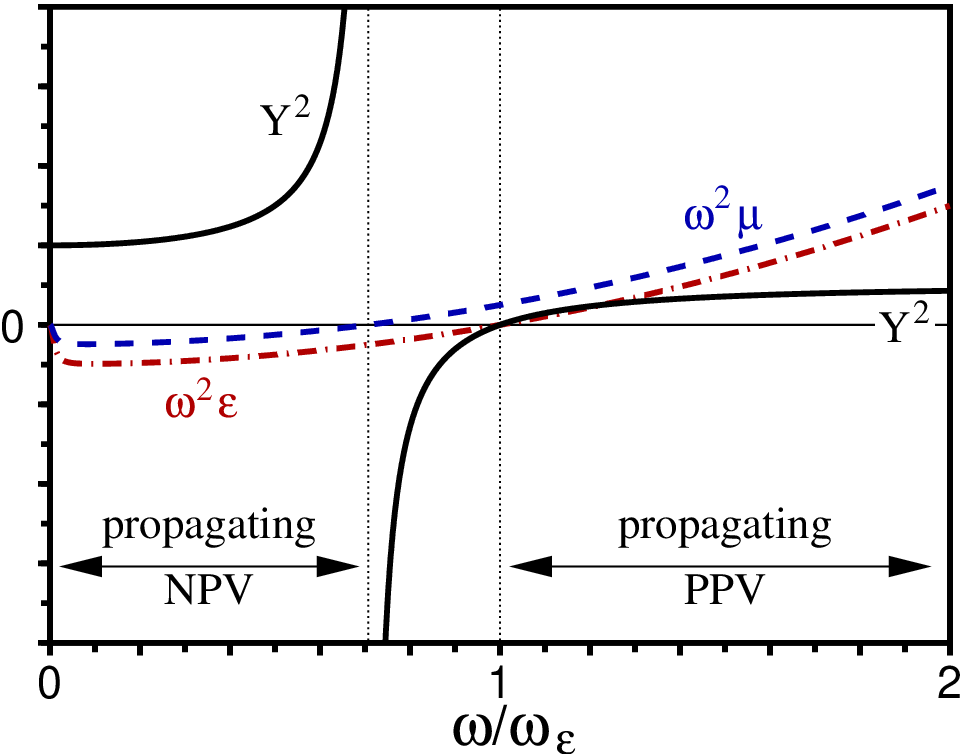}
\caption{
Normalised electromagnetic field ratio and $\epsilon,\mu$ curves for 
 Drude responses in both permittivity $\epsilon_r$
 and permeability $\mu_r$.
The magnetic resonance at $\omega_\mu = \omega_\epsilon/\sqrt{2}$
 is lower than 
 the (dielectric) plasma frequency $\omega_\epsilon$, 
 and $\gamma_\epsilon=\gamma_\mu=0.01 \omega_\epsilon$.
Large $|Y|$ corresponds to mostly magnetic $\Vec{G}^\pm$ fields, 
 small $|Y|$ corresponds to mostly electric $\Vec{G}^\pm$.
The frequency ranges of propagating negative phase velocity (NPV)
 and positive phase velocity (PPV) light are shown.
}
\label{F-DD}
\end{figure}

\begin{figure}
 \includegraphics[width=0.70\columnwidth,angle=0]{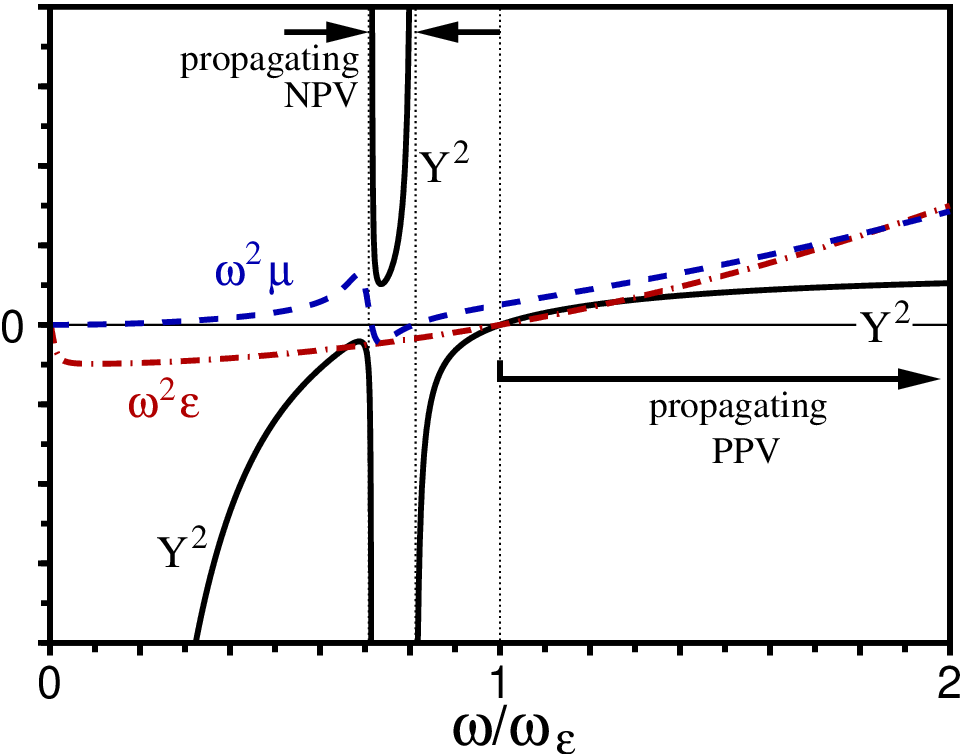}
\caption{
Normalised electromagnetic field ratio and $\epsilon,\mu$ curves for 
 Drude response permittivity $\epsilon_r$ (dot-dashed line), 
 with a pseudo-Lorentz response for the permeability $\mu_r$ (dashed line).
The magnetic resonance at $\omega_\mu = \omega_\epsilon/\sqrt{2}$
 is lower than 
 the (dielectric) plasma frequency $\omega_\epsilon$, 
 and $\gamma_\epsilon=\gamma_\mu/5=0.01 \omega_\epsilon$, 
 $F_\mu=5$.
The high-frequency behaviour of the pseudo-Lorentz model 
 ``illegally'' increases faster than that of the Drude model, 
 whereas a properly causal form should match it.
Apart from detail, 
 and the high-frequency behaviour, 
 replacing the pseudo-Lorentz model
 with the (causal) Lorentz one gives a figure of similar appearance.
The frequency ranges of propagating NPV and PPV light are shown.
}
\label{F-DF}
\end{figure}

Figures \ref{F-DD} and \ref{F-DF} show how $Y$
 varies with frequency for two different metamaterial types,
 with the dispersions encoded on $\epsilon_r$ and $\mu_r$
 and scaled by $\omega^2$
 to moderate the low-frequency singularity of the Drude response.
The extreme limits of large $Y$ occur when $|\mu_r| \ll |\epsilon_r|$,
 that is,
 usually just at an edge of a non-propagating band, 
  where $\mu_r$ is about to change sign.
In such a region, 
 it would be better to revert to the $\Vec{G}^\pm$ fields, 
 or to rescale the propagation equations into units of magnetic field
 (e.g., with some $\Vec{K}^\pm = \Vec{G}^\pm / 2 \beta_r$).

\begin{figure}
 \includegraphics[width=0.70\columnwidth,angle=0]{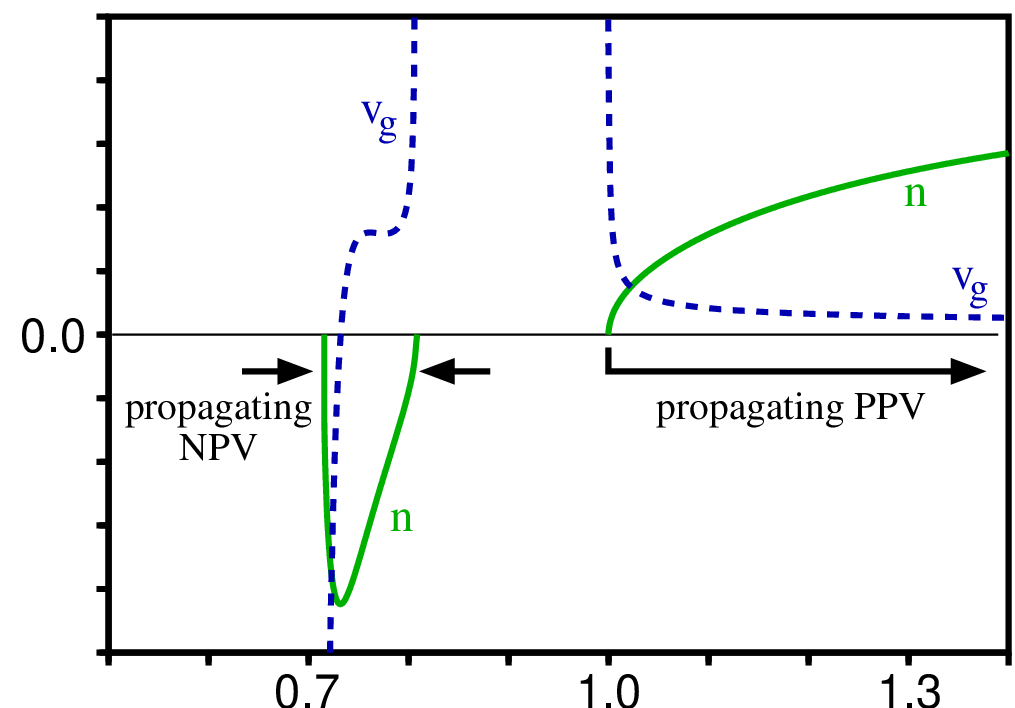}
\caption{
Normalised refractive index $n$ (solid line)
 and group velocity $v_g=dk/d\omega$ (dashed line)
 for the system defined in fig. \ref{F-DF}, 
 for only those frequencies where the field propagates.
By comparing with fig. \ref{F-DF}, 
 we can see that the edges of the NPV band are dominated
 by the magnetic field, 
 whereas the lower edge of the PPV band is dominated
 by the electric field.
Note that in the NPV band, 
 the sign of the group velocity is not tied
 to the sign of the phase velocity.
}
\label{F-DF-n}
\end{figure}

In previous work 
 \cite{Scalora-SAPDMBZ-2005prl,Wen-XDTSF-2007pra,DAguanno-MB-2008josab},
 a Drude type response for both $\epsilon$ and $\mu$ was assumed, 
 where $\epsilon_r, \mu_r \propto 1 - \omega_{\epsilon,\mu}^2 
                         / (\omega^2 - \imath \gamma_{\epsilon,\mu} \omega)$; 
 and this situation is shown on Fig. \ref{F-DD}.
However, 
 although the dielectric response in metamaterials
 (e.g., a wire grid array \cite{Pendry-HSY-1996prl,Pendry-HRS-1998jpcm})
 often has this behaviour, 
 the magnetic response of split ring resonators (SRRs) differs.
SRR magnetization is best described by a pseudo-Lorentz model\footnote{Also
   known as the ``F-model''}
 \cite{Pendry-HRS-1999ieeemtt} with 
 $\mu_r \propto 1 + F_\mu \omega^2 
                      / (\omega^2 - \omega_\mu^2 - \imath \gamma_\mu \omega)$,
 although sometimes a true Lorentz response is used instead, 
 $\mu_r \propto 1 + F_\mu \omega_\mu^2 
                      / (\omega^2 - \omega_\mu^2 - \imath \gamma_\mu \omega)$.
Note the difference between the numerators in these latter two expressions:
 a frequency-dependent $\omega^2$ 
 versus a constant material parameter $\omega_\mu^2$.
The pseudo-Lorentz model
 has an incorrect high-frequency behaviour, 
 and so it is incompatible with the Kramers-Kronig relations
 that enforce causality.
However,  
 at low and medium frequency
 it is a better match to the physical response of SRRs, 
 and so I use it for Figs. \ref{F-DF} and \ref{F-DF-n}.

There are frequency ranges over which the linear material responses 
 vary dramatically, 
 and in particular on Fig. \ref{F-DF-n}
 (which uses the same model as Fig. \ref{F-DF})
 this is evident for both the refractive index $n$
 and group velocity $v_g$.
If we aim to operate in such regions, 
 this leads to two potential complications.
First, 
 if we have chosen reference parameters that do not match 
 the linear material responses exactly, 
 then the correction terms will become large, 
 meaning that our wavelength-scale ``slow evolution'' approximation
 may come under threat.
Second, 
 even if our reference parameters do match
 the linear material responses exactly, 
 our wavelength-scale will have become frequency dependent, 
 and so again our ``slow evolution'' approximation may be threatened.
In either case the solution is simple -- 
 we just need to revert to the bi-directional wave equations 
 [i.e., eqns. \eqref{eqn-1d-dzGxpm-raw} or \eqref{eqn-1d-dzGxpm}].
However,
 this does not necessarily mean 
 that any backward evolving fields are generated
 (as explained in \cite{Kinsler-RN-2005pra}, 
  and following a different approach in \cite{Kinsler-2008-fchhg}), 
 so that in principle one could optimize the propagation
 by reinstating it only over those frequency ranges
 where it becomes necessary.

%
\section{Conclusions}\label{S-conclude}

I have derived a 
 uni-directional optical pulse propagation equation
 for media with both electric and magnetic responses, 
 based on the directional fields approach \cite{Kinsler-RN-2005pra}.
This involved a
 re-expression of Maxwell's equations, 
 and required only a single approximation to reduce
 a one dimensional bi-directional model,
 to a uni-directional first-order wave equation.
The simplicity of this approach makes it very 
 convenient in waveguides,
 optical fibres, 
 or other collinear situations.
The important approximation is that the pulse
 evolves only slowly on the scale of a wavelength; 
 and indeed this is a valid assumption
 in a wide variety of cases -- 
 note in particular that nonlinear effects
 have to be unrealizably strong to violate it \cite{Kinsler-2007josab}.
The result has no intrinsic bandwidth restrictions,
 makes no demands on the pulse profile, 
 and does not require a co-moving frame --
 unlike other common types of derivation
 \cite{Agrawal-NFO,Brabec-K-1997prl,Geissler-TSSKB-1999prl,Kinsler-N-2003pra}.

The resulting equations have the advantage
 that they are straightforward to write down, 
 despite containing the complications of \emph{both}
 electric and magnetic responses, 
 and that a carrier-envelope representation
 or co-moving frames are easy to apply if desired, 
 requiring no further approximation.
In this, 
 they match the clarity and flexibility 
 of factorized second-order wave equations
 \cite{Ferrando-ZCBM-2005pre,Genty-KKD-2007oe,Kinsler-2008-fchhg}, 
 but they can more easily incorporate
 the effects of magnetic material responses --
 albiet 
 at the cost of being restricted to one dimensional propagation.

%
\acknowledgments

I acknowledge financial support from the
 Engineering and Physical Sciences Research Council
 (EP/E031463/1).


\bibliography{/home/physics/_work/bibtex.bib}

\appendix

\section{Derivation of eqn. \protect{\eqref{EQN-BASIC-DZGPM}}}\label{S-derive}

The derivation in this paper is simpler, more general, 
 and defines $\Vec{G}^\pm$ using a better sign convention
 than those in \cite{Kinsler-RN-2005pra,Kinsler-2006arXiv-fleck}.
I start with the Maxwell curl equations, 
 and transform into frequency space:
~
\begin{align}
\grad \cross \Vec{H} (t)
&=
  +\partial_t \epsilon_r \convol \Vec{E} (t)
  + \Vec{J} (t)
,
\nonumber
\\
\grad \cross \Vec{E} (t)
&=
  -\partial_t  \mu_r \convol \Vec{H} (t)
  -\mu_0 \Vec{K} (t)
\\
\grad \cross \Vec{H} (\omega)
&=
  - \imath \omega 
    ~ \alpha_r (\omega) ^2  
    \Vec{E} (\omega)
  + \Vec{J} (\omega)
,
\nonumber
\\
\grad \cross \Vec{E} (\omega)
&=
  + \imath \omega  
    ~ \beta_r (\omega)^2  
    \Vec{H} (\omega)
  -
   \mu_0 \Vec{K} (\omega)
.
\end{align}
I now rotate the $\grad \cross \Vec{H}$ equation 
 by taking the cross product with $\Vec{u}$,
\begin{align}
~
\Vec{u} \cross \left( \grad \cross \Vec{H} \right)
&=
  - \imath \omega   
    ~\alpha_r^2
    \left(
      \Vec{u} ~ \cross \Vec{E}
    \right)
  + \Vec{u} \cross \Vec{J}
,
\label{eqn-u-curl-H}
\end{align}
scale each part by $\beta_r$ and $\alpha_r$ respectively, 
 while insisting that these parameters do not depend on position.
Thus,
~
\begin{align}
\Vec{u} \cross \left( \grad \cross \beta_r \Vec{H} \right)
&=
  - \imath \omega   
    ~\beta_r \alpha_r^2   
    \left(
      \Vec{u} ~ \cross \Vec{E}
    \right)
  + \Vec{u} \cross  \beta_r \Vec{J}
,
\nonumber
\\
\grad \cross \alpha_r \Vec{E}
&=
   + \imath \omega   
    ~\alpha_r 
    \beta_r^2   
    \Vec{H}
 -
  \alpha_r 
  \mu_0 \Vec{K}
,
\end{align}
and then take the sum and difference --
~
\begin{align}
  \grad \cross \alpha_r \Vec{E}
&
 \pm
 ~
  \Vec{u} \cross \left( \grad \cross \beta_r \Vec{H} \right)
\nonumber\\
&=
 +
  \imath \omega   
    ~\alpha_r 
    \beta_r^2
    \Vec{H}
 ~
 \mp 
 ~
  \imath \omega   
    ~\beta_r 
    \alpha_r^2   
    \left(
      \Vec{u} ~ \cross \Vec{E}
    \right)
 ~~
\nonumber
\\
&
 \qquad \qquad \qquad 
 \pm 
 ~~
  \Vec{u} \cross  \beta_r \Vec{J}
 -
  \alpha_r
  \mu_0
  \Vec{K}
.
\end{align}

The vector $\Vec{G}^\pm$ fields are defined in eqn. \eqref{eqn-basic-Gpm}, 
 but that neglects the longitudinal part of $\Vec{H}$.
Thus, 
 for completeness, 
 we also need to define ${G}^\circ = \Vec{u} \cdot \beta_r \Vec{H}$
 (as in \cite{Kinsler-RN-2005pra,Kinsler-2006arXiv-fleck}).
Their form means that I need to convert both 
 the second term on the LHS of the sum-and-difference equation above, 
 as well as the RHS.  
It is most important for the LHS to be simple, 
 because this will define the type of propagation specified by the RHS.  
In the following I use the vector identity.
~
\begin{align}
  \Vec{u} \cross \left( \grad \cross \Vec{H} \right)
 -
  \grad \left( \Vec{u} \cdot \Vec{H} \right)
&=&
  \grad \cross \left( \Vec{u} \cross \Vec{H} \right)
,
\end{align}
along with 
 $\Vec{u} \cross [\Vec{u} \cross \Vec{H}]
 = [ \Vec{u} \cdot \Vec{H} ] \Vec{u} - \Vec{H} $,
 so that 
~
\begin{align}
  \Vec{u} \cross \Vec{G}^{\mp}
&=
  \Vec{u} \cross \alpha_r \Vec{E}
 \pm
  \Vec{u} \cross 
    \left[ 
      \Vec{u} \cross \beta_r \Vec{H} 
    \right]
\\
&=
  \Vec{u} \cross \alpha_r \Vec{E}
 \mp
  \beta_r \Vec{H} 
 ~~
 \pm
  \left[ \Vec{u} \cdot \beta_r \Vec{H} \right] \Vec{u} 
\\
&=
  \Vec{u} \cross \alpha_r \Vec{E}
 \mp
  \beta_r \Vec{H} 
 ~~
 \pm
  \Vec{u} {G}^\circ
.
\end{align}

\begin{widetext}
Continuing the derivation, 
~
\begin{align}
  \grad \cross \alpha_r \Vec{E}
 ~~
 \pm
 ~~
  \Vec{u} \cross \left( \grad \cross \beta_r \Vec{H} \right)
&=
 +
  \imath \omega \alpha_r \beta_r^2 
  \Vec{H}
 ~~
 \mp 
 ~~
  \imath \omega \beta_r \alpha_r^2 
  \left( \Vec{u} \cross \Vec{E} \right)
 ~~
 \pm 
  \Vec{u} \cross  \beta_r \Vec{J}
 -
  \alpha_r 
  \mu_0 \Vec{K}
,
\\
  \grad \cross \alpha_r \Vec{E}
 ~~
 \pm
 ~~
  \grad \cross \left( \Vec{u} \cross \beta_r \Vec{H} \right)
 \pm
  \grad \left( \Vec{u} \cdot \beta_r \Vec{H} \right)
&=
 +
  \imath \omega \alpha_r \beta_r^2 
  \Vec{H}
 ~~
 \mp 
 ~~
  \imath \omega \beta_r \alpha_r^2 
  \left( \Vec{u} \cross \Vec{E} \right)
 ~~
 \pm 
  \Vec{u} \cross  \beta_r \Vec{J}
 -
  \alpha_r 
  \mu_0 \Vec{K}
\\
  \grad \cross 
  \left[
    \alpha_r \Vec{E}
    ~~
    \pm
    ~~
    \left( \Vec{u} \cross \beta_r \Vec{H} \right)
  \right]
&=
 +
  \imath \omega \alpha_r \beta_r^2 
  \Vec{H}
 ~~
 \mp 
 ~~
  \imath \omega \beta_r \alpha_r^2 
  \left( \Vec{u} \cross \Vec{E} \right)
 \mp
  \grad \left( \Vec{u} \cdot \beta_r \Vec{H} \right)
 ~~
 \pm 
  \Vec{u} \cross  \beta_r \Vec{J}
 -
  \alpha_r 
  \mu_0 \Vec{K}
~~~~ ~~~~
\\
  \grad \cross \Vec{G}^{\mp}
&=
  \imath \omega 
  \left\{
    \alpha_r \beta_r^2 
    \Vec{H}
   ~~
   \mp 
   ~~
    \beta_r \alpha_r^2 
    \left( \Vec{u} \cross \Vec{E} \right)
  \right\}
 \mp
  \grad \left( \Vec{u} \cdot \beta_r \Vec{H} \right)
 ~~
 \pm 
  \Vec{u} \cross  \beta_r \Vec{J}
 -
  \alpha_r 
  \mu_0 \Vec{K}
\label{eqn-Gcurl-mid}
\end{align}

I now rearrange eqn. \eqref{eqn-Gcurl-mid} to give the final form,
 in which I substitute $\partial_t \Vec{P} = \Vec{J}(t)$
 and $\partial_t \Vec{M} = \Vec{K}(t)$ to match eqn. \eqref{EQN-BASIC-DZGPM}.
Thus
~
\begin{align}
  \grad \cross \Vec{G}^{\mp}
&=
  \imath \omega 
  \left\{
    \alpha_r \beta_r
    \Vec{u}
    {G}^\circ
   ~~
   -
    \alpha_r \beta_r^2 
    \left(
      \Vec{u} \cross \left[ \Vec{u} \cross \Vec{H} \right]
    \right)
   ~~
   \mp 
    \beta_r \alpha_r^2 
    \left(
      \Vec{u} \cross \Vec{E} 
    \right)
  \right\}
 \mp
  \grad {G}^\circ
 ~~
 \pm 
  \Vec{u} \cross  \beta_r \Vec{J}
 -
  \alpha_r 
  \mu_0
  \Vec{K}
\\
&=
 \mp 
  \imath \omega 
  \left\{
    \beta_r \alpha_r^2 
    \left(
      \Vec{u} \cross \Vec{E}
    \right)
   \pm
    \alpha_r \beta_r^2 
    \left(
      \Vec{u} \cross \left[ \Vec{u} \cross \Vec{H} \right]
    \right)
  \right\}
 ~~
 +
  \imath \omega 
    \alpha_r \beta_r
    \Vec{u} {G}^\circ
 ~~
 \mp
  \grad {G}^\circ
 ~~
 \pm 
  \Vec{u} \cross  \beta_r \Vec{J}
 -
  \alpha_r 
  \mu_0
  \Vec{K}
,
\end{align}
and finally
\begin{align}
  \grad
 \cross
  \Vec{G}^\pm
&=
 \pm
  \imath \omega ~
    \alpha_r \beta_r
  \Vec{u}
  \cross
  \Vec{G}^\pm
 ~~
 +
  \imath \omega 
    \alpha_r \beta_r
    \Vec{u} {G}^\circ
 ~~
 \pm
  \grad {G}^\circ
 ~~
 \mp 
  \Vec{u} \cross  \beta_r \Vec{J}
 -
  \alpha_r 
  \mu_0
  \Vec{K}
\\
&=
 \pm
  \imath \omega ~
    \alpha_r \beta_r
  \Vec{u}
  \cross
  \Vec{G}^\pm
 ~~
 +
  \imath \omega 
    \alpha_r \beta_r
    \Vec{u} {G}^\circ
 ~~
 \pm
  \grad {G}^\circ
 ~~
 \pm
  \imath \omega \beta_r 
  \Vec{u} \cross \Vec{P}
 +
  \imath \omega \alpha_r 
  \mu_0
  \Vec{M}
.
\end{align}
\end{widetext}

Note that for $\Vec{J}(t) = \partial_t \Vec{P} 
                          = \partial_t \kappa_\epsilon \Vec{E}(t)$, 
 where $\kappa_\epsilon$ is some complicated but scalar
 dielectric response function,
 we have
~
\begin{align}
 \mp
  \Vec{u} 
 \cross
  \beta_r \Vec{J}(\omega)
&=
 \pm
  \imath \omega \beta_r 
  \Vec{u} 
 \cross
  \Vec{P}
\quad
=
 \pm
  \imath \omega 
  \alpha_r^2
  \beta_r 
  \kappa_\epsilon
  \convol
  \Vec{u} 
 \cross
  \Vec{E}
\\
&=
 \pm
  \imath \omega 
  \frac{\alpha_r^2\beta_r}
       {2}
  \kappa_\epsilon
  \convol
  \left(
    \Vec{u} \cross 
    \left[ \Vec{G}^+ + \Vec{G}^- \right]
  \right)
,
\end{align}
and for $\Vec{K}(t) = \partial_t \Vec{M} 
                    = \partial_t \kappa_\mu \Vec{H}(t)$, 
 where $\kappa_\mu$ is some complicated but scalar
 magnetic response function,
 we have
~
\begin{align}
 -
  \alpha_r \mu_0 \Vec{K}(\omega)
&=
 +
  \imath \omega 
  \mu_0
  \alpha_r
  \Vec{M}
\quad
=
 +
  \imath \omega 
  \mu_0
  \alpha_r
  \kappa_\mu
  \convol
  \Vec{H}
\\
&=
 +
  \imath \omega 
  \mu_0
  \alpha_r
  \kappa_\mu
  \convol
  \left(
    \Vec{u}
    \left[
      \Vec{u} \cdot \Vec{H}
    \right]
   -
    \Vec{u} 
    \cross
    \Vec{u} 
    \cross
    \Vec{H}
  \right)
\\
&=
 +
  \imath \omega 
  \mu_0
  \frac{\alpha_r}{\beta_r}
  \kappa_\mu
  \convol
  \left(
    \Vec{u} {G}^\circ
   -
    \Vec{u} \cross 
    \left[
      \Vec{u} \cross \beta_r \Vec{H}
    \right]
  \right)
\\
&=
 -
  \imath \omega 
  \mu_0
  \frac{\alpha_r}{2\beta_r}
  \kappa_\mu
  \convol
  \left(
    \Vec{u} \cross 
    \left[
      \Vec{G}^+ - \Vec{G}^-
    \right]
   -
    2 \Vec{u} {G}^\circ
  \right)
.
\end{align}

Finally, 
 when generating eqn. \eqref{eqn-u-curl-H}, 
 we lost the longitudinal part of 
 $\grad \cross \Vec{H} = \partial \epsilon_r \Vec{E} + \Vec{J}$ 
 (i.e. that parallel to $\Vec{u}$).
This is
~
\begin{align}
  \Vec{u}
 \cdot
  \grad \cross \Vec{H}
&=
 -
  \imath \omega \alpha_r^2
  \Vec{u}
 \cdot
  \Vec{E}
 +
  \Vec{u}
 \cdot
  \Vec{J}
\\
  \Vec{u}
 \cdot
  \grad \cross 
  \left(
    \Vec{G}^+ - \Vec{G}^- + 2 \Vec{u} G^\circ
  \right)
&=
 -
  \imath \omega \alpha_r \beta_r
  \Vec{u}
 \cdot
  \left(
    \Vec{G}^+ - \Vec{G}^-
  \right)
\nonumber
\\
& \qquad
 +
  2 \beta_r
  \Vec{u}
 \cdot
  \Vec{J}
\\
  2
  \Vec{u} \cdot 
  \left(
    \grad {G}^\circ
   -
    \Vec{u} \cross \grad {G}^\circ
  \right)
&=
 -
  \imath \omega \alpha_r \beta_r
  \Vec{u}
 \cdot
  \left(
    \Vec{G}^+ - \Vec{G}^-
  \right)
\nonumber
\\
& \qquad
 +
  2 \beta_r
  \Vec{u}
 \cdot
  \Vec{J}
\end{align}
\begin{align}
  2
  \Vec{u} \cdot 
    \grad {G}^\circ
&=
 -
  \imath \omega \alpha_r \beta_r
  \Vec{u}
 \cdot
  \left(
    \Vec{G}^+ - \Vec{G}^-
  \right)
 +
  2 \beta_r
  \Vec{u}
 \cdot
  \Vec{J}
,
\end{align}
 since
 $\grad \cross {G}^\circ \Vec{u} 
    = {G}^\circ \grad \cross \Vec{u} 
      - \Vec{u} \cross \grad {G}^\circ
    = - \Vec{u} \cross \grad {G}^\circ$,
 and
~
\begin{align}
  \Vec{u} \cdot
  \grad \cross 
  \left(
    \Vec{G}^+ - \Vec{G}^-
  \right)
&=
  \imath \omega
  \alpha_r \beta_r
  \Vec{u} \cdot
  \Vec{u} \cross 
  \left(
    \Vec{G}^+ + \Vec{G}^-
  \right)
\nonumber
\\
& \qquad
 +
  2 
  \Vec{u} \cdot
  \grad {G}^\circ
 +
  2 \imath \omega \beta_r
  \Vec{u} \cdot
   \Vec{u} \cross \Vec{P}
\\
&=
  2 
  \Vec{u} \cdot
  \grad {G}^\circ
.
\end{align}

%
\section{Correction terms}\label{S-corrections}

In this appendix I work through the details
 of how the polarization and magnetization terms 
 scale with respect to one another.
To simplify matters, 
 I assume all corrections are scalar
 since when $\epsilon_r$ and $\mu_r$ are not field-polarization 
 or orientation sensitive, 
 the scalings remain the same, 
 even if the specific field terms may vary
 (e.g. ${E}_x {E}_y$ instead of ${E}_x^2$).

Consider the general unidirectional equation 
 for ${F}_x^\pm$ (i.e. eqn. \eqref{eqn-1d-dzGxpm}),
 and replace the polarization and magnetization terms
 with dimensionless response parameters $q_\epsilon$ and $q_\mu$
 multiplied by the appropriate field ${E}_x$ or ${H}_y$.
Then replace ${E}_x$ and ${H}_y$
 with their representation in terms of ${F}_x^+$,
 so that 
~
\begin{align}
  \frac{\imath \omega}
       {2}
  \frac{\beta_r}
       {\alpha_r}
  {P}_x
 +
  \frac{\imath \omega}
       {2}
  \mu_0 
  {M}_y
&=
  \frac{\imath \omega}
       {2}
  \frac{\beta_r}
       {\alpha_r}
  q_\epsilon
  \epsilon_0
  {E}_x
 ~~
 +
  \frac{\imath \omega}
       {2}
  q_\mu
  \mu_0 
  {H}_y
\label{eqn-corrections-generalEH}
\\
&=
  \frac{\imath \omega}
       {2}
 \left[
  \frac{\beta_r}
       {\alpha_r}
  q_\epsilon
  \epsilon_0
  {F}_x^+
 ~~
 +
  q_\mu
  \mu_0 
  \frac{\alpha_r}
       {\beta_r}
  {F}_x^+
 \right]
\\
&=
  \frac{\imath \omega \alpha_r \beta_r}
       {2}
 \left[
  \frac{q_\epsilon \epsilon_0}
       {\alpha_r^2}
  {F}_x^+
 ~~
 +
  \frac{q_\mu \mu_0}
       {\beta_r^2}
  {F}_x^+
 \right]
\\
&=
  \frac{\imath k_r}
       {2}
  \frac{\epsilon_0}
       {\epsilon_r}
 \left[
  q_\epsilon
 ~~
 +
  q_\mu
  \frac{\mu_0}
       {\epsilon_0}
  \frac{\epsilon_r}
       {\mu_r}
 \right]
  {F}_x^+
,
\label{eqn-corrections-generalFp}
\end{align}
 remembering that $\Vec{F}^+ = \Vec{E} = (\beta_r/\alpha_r) \Vec{H}$, 
 and that $\epsilon_r=\alpha_r^2$, 
 and $\mu_r=\beta_r^2$.

Since we consider the electric-field-like field ${F}^+$, 
 the polarization corrections are trivial to write down;
 as for an $m$-th order nonlinear term,
 $q_\epsilon = \chi_\epsilon F^{+(m-1)}$.
This means we need only concentrate on the magnetization correction.
If $q_\mu$ is that for an $m$-th order nonlinear term, 
 then $q_\mu = \chi_\mu H^{n-1} 
             = \chi_\mu (\alpha_r/\beta_r)^{(m-1)} F^{+(m-1)}$.
Writing down only the term in square brackets from 
 eqn. \eqref{eqn-corrections-generalFp}
 gives us
~
\begin{align}
&
 \left[
  q_\epsilon
 ~
 +
  \chi_\mu
  \frac{\mu_0}
       {\epsilon_0}
  \frac{\epsilon_r}
       {\mu_r}
  \left(
    \frac{\alpha_r}
         {\beta_r}
  \right)^{m-1}
  {F}_x^{+(m-1)}
 \right]
  {F}_x^+
\\
&\qquad \qquad =
 \left[
  q_\epsilon
 ~
 +
  \chi_\mu
  \frac{\mu_0}
       {\epsilon_0}
  \left(
  \frac{\epsilon_r}
       {\mu_r}
  \right)^{\frac{m+1}{2}}
  {F}_x^{+(m-1)}
 \right]
  {F}_x^+
.
\end{align}
Note that corrections for linear loss or gain
 are first-order processes (i.e. with $m=1$),
 where for loss we need $q \sim \imath \gamma$, 
 with $\gamma>0$; 
Thus for loss the whole correction term will be proportional 
 to $-\gamma {F}_x^+$, 
 as would be expected.

\end{document}